\documentclass{pasj00}
\draft

\begin{document}
\SetRunningHead{K. Haze et al.}{Experimental demonstration of binary shaped pupil mask coronagraphs for telescopes with obscured pupils}

\title{Experimental demonstration of binary shaped pupil mask coronagraphs for telescopes with obscured pupils}

\author{
   Kanae \textsc{Haze}\altaffilmark{1}
   Keigo \textsc{Enya}\altaffilmark{1}
   Lyu \textsc{Abe}\altaffilmark{2}
   Aoi \textsc{Takahashi}\altaffilmark{1,3}
   Takayuki \textsc{Kotani}\altaffilmark{4}
   and
   Tomoyasu \textsc{Yamamuro}\altaffilmark{5}}
 \altaffiltext{1}{Institute of Space and Astronautical Science, Japan Aerospace Exploration Agency (JAXA), 
        3-1-1 Yoshinodai, Chuo-ku, Sagamihara, Kanagawa 252-5210, Japan}
        \email{haze@ir.isas.jaxa.jp}
 \altaffiltext{2}{Laboratoire Hippolyte Fizeau,
        UMR 6525 Universit\'e de Nice-Sophia Antipolis,
        Parc Valrose, F-06108 Nice, France}
 \altaffiltext{3}{Department of Space and Astronautical Science,
        The Graduate University for Advanced Studies, 3-1-1 Yoshinodai, Chuo-ku, Sagamihara, Kanagawa 252-5210, Japan}
 \altaffiltext{4}{Inter-University Research Institute Corporation, National Institutes of Natural Sciences, National Astronomical Observatory of Japan, 2-21-1 Osawa, Mitaka,Tokyo 181-8588, Japan}\altaffiltext{5}{Optcraft, 3-26-8 Aihara, Sagamihara, Kanagawa 229-1101, Japan}

%


\maketitle

\begin{abstract}
We present the fabrication and experimental demonstration of three free-standing
binary shaped pupil mask coronagraphs, which are applicable for telescopes with partially
obscured pupils. Three masks, designed to be complementary (labeled Mask-A,
Mask-B, and Mask-C), were formed in 5 $\micron$ thick nickel. The design of Mask-A is based
on a one-dimensional barcode mask. The design principle of Mask-B is similar, but has
a smaller inner working angle and a lower contrast than Mask-A. Mask-C is based on a
concentric ring mask and provides the widest dark region and a symmetric point spread
function. Mask-A and Mask-C were both designed to produce a flexibly tailored dark
region (i.e., non-uniform contrast). The contrast was evaluated using a light source comprising
a broadband super-luminescent light-emitting diode with a center wavelength
of 650 nm, and the measurements were carried out in a large vacuum chamber. Active
wavefront control was not applied in this work. The coronagraphic images obtained by
experiment were mostly consistent with the designs. 
The contrast of Mask-A within the ranges 3.3 - 8$\lambda/D$ and 8 - 12$\lambda/D$ was $\sim10^{-4}$ - $10^{-7}$ and $\sim10^{-7}$, respectively, where $\lambda$ is the wavelength and $D$ is the pupil diameter. 
The contrast close to the center of Mask-B was $\sim10^{-4}$ and that of Mask-C over an extended field of view (5 - 25 $\lambda/D$) was $\sim10^{-5}$ - $10^{-6}$. 
The effect of tilting the masks was investigated, and found to be irrelevant at the $\sim10^{-7}$  contrast level. Therefore the masks can be tilted to avoid ghosting. These high-contrast free-standing masks have the potential to enable coronagraphic imaging over a wide wavelength range using
both ground-based and space-borne general-purpose telescopes with pupil structures
not specifically designed for coronagraphy.

\end{abstract}

\section{Introduction}
Direct observation of spatially resolved exoplanets is important
in order to learn about their formation, evolution,
and diversity. However, the huge contrast in flux between
an exoplanet and its parent star is the primary difficulty
in accomplishing direct observations. For instance, the
contrast between the sun and the earth is $\sim10^{-10}$ in the
visible wavelength region, and $\sim10^{-6}$ in the mid-infrared
(mid-IR) wavelength region \citep{Traub}.
Therefore, it is necessary to develop stellar coronagraphs
that can overcome the contrast between the star and the
planet.Among the various coronagraphic imaging methods,
binary shaped pupil mask coronagraphs have some advantages
\citep{Jacquinot, Spergel, Vanderbei2003a, Vanderbei2003b, Vanderbei2004, Kasdin2005a, Kasdin2005b, Tanaka, Belikov2007, Enya2007, Enya2008, Enya2011b, Haze2009, Haza2011, EnyaAbe2010, Carlotti, Haze2012}. 
The function of a binary shaped pupil mask coronagraph is to produce a
high-contrast point spread function (PSF) which is much
less sensitive to telescope pointing errors, and also less sensitive
to wavelength (exceptwhen scaling the size of the PSF)
than other coronagraphs. Simplicity is another advantage
of this type of optical system.
In our previous experimental research \citep{Enya2007, Enya2008, Haze2009, Haze2011}, we demonstrated
high contrast performance using a checkerboard mask,
which is a conventional type of binary shaped pupil mask
\citep{Vanderbei2004}. After the successful demonstration
of the principle of binary shaped pupil masks fabricated
on substrates, free-standing metal (copper and nickel)
masks were developed. Since there is no substrate limiting
the applicable wavelength region of the mask, such freestanding
masks can be used for observations over a wide
infrared wavelength region. This provides a big advantage,
since the contrast between the star and the planet
is much less in the infrared region than in the visible light
region. Moreover, high contrasts ($\sim10^{-7}$) have been confirmed
in experiments using free-standing masks at visible
wavelengths \citep{Haze2012, Enya2012}.
The mask design used in those experiments, a conventional
checkerboard design, is applicable to specifically
designed off-axis telescopes. However, the high contrast
performance of these masks is impaired when used with a
centrally obscured telescope and in which there is partial
obscuration due to support arms. In this case, further optimization
is required. \cite{EnyaAbe2010} presented solutions
using a simple one dimensional (1D) optimization of
the mask design so that the masks could be applied to various
telescope pupils including those with segments and/or
obscurations. More advanced, complicated mask designs
for telescopes with obscured pupils are presented in \cite{Enya2011b, Carlotti}.
Considering this background, we carried out the fabrication
and experimental demonstration of free-standing mask
coronagraphs that can be used with centrally obscured onaxis
telescopes. The experiments, results, and a discussion
are presented in the following sections.

\section{Experiments and Results}
In this section, we present three new designs for free-standing masks and the results of our first coronagraphic experiments using each mask. 

\subsection{Mask designs and fabrication}
\begin{figure}[hbtp]
                                      \begin{center}
                                       \includegraphics[width=14cm]{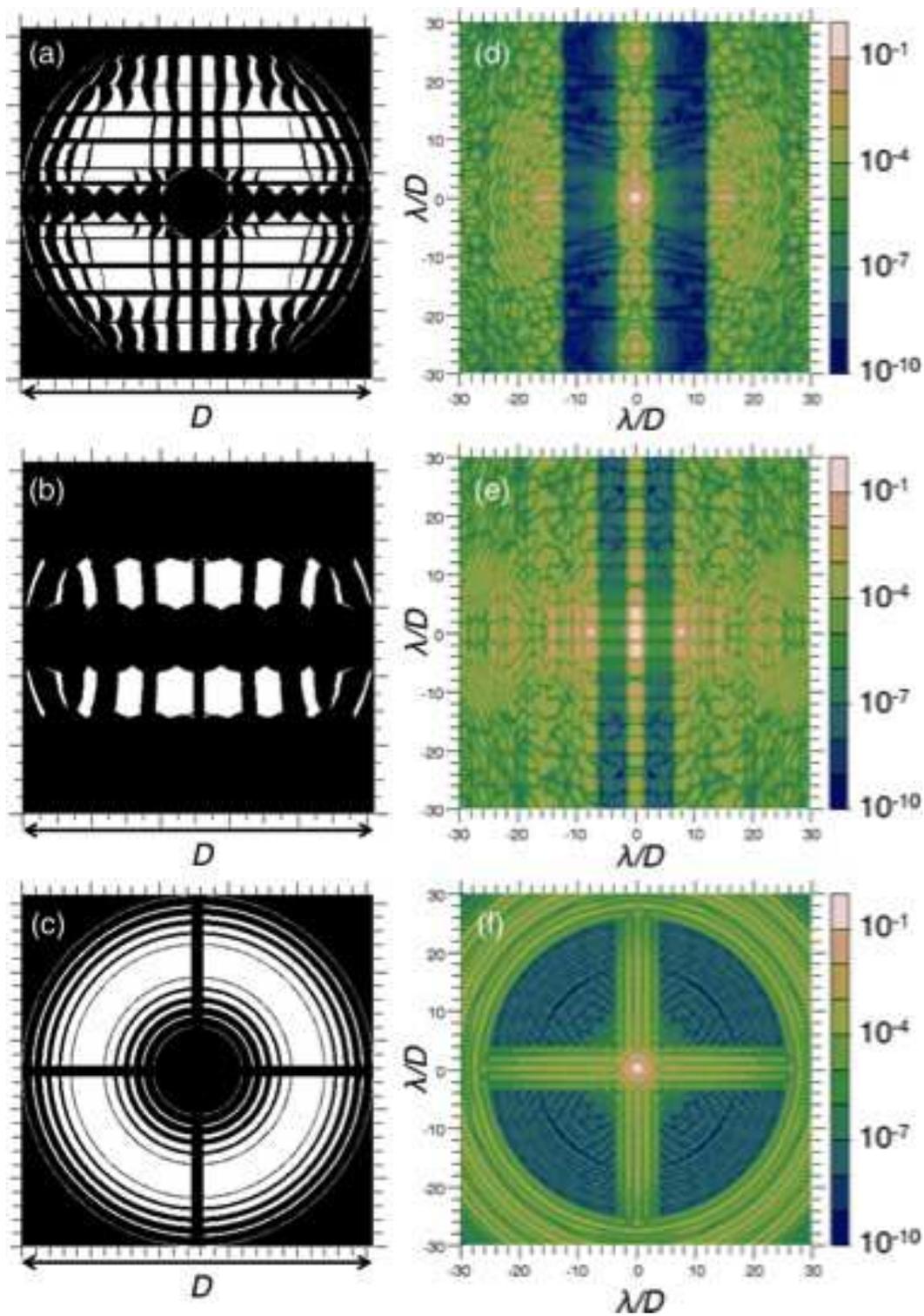}
                                      \end{center}
                                      \caption{Left: the pupil mask designs: (a) Mask-A, (b) Mask-B, and (c) Mask-C. 
                                      The transmission through the black and white regions is 0 and 1, respectively. 
                                      The diameter of the circle encompassing the transmissive region is 10mm. 
                                      Right: the expected (theoretical) PSFs for the pupil masks.
                                      }
                                      \label{fig1}
                                     \end{figure}

The designs of the masks (Mask-A, Mask-B, and Mask-C)
are shown on the left-hand side of figure \ref{fig1}. These masks
are types of binary shaped pupil masks, and their basic
design is based on those reported in \cite{Enya2011b}.
Optimization of the basic mask shapes was performed with
the LOQO solver presented by \cite{Vanderbei1999}. Mask-A
and Mask-B are based on an integral 1D design, and
Mask-C is based on a concentric ring design (table \ref{table1}).
These masks have the general advantages of binary pupil
masks: (1) They are robust against pointing errors, and
(2) they can, in principle, make observations over a wide
range of wavelengths. These three masks also have a
particularly important asset: (3) the design makes them
applicable for the pupil of SPICA telescope, which is
partially obscured by a secondary mirror and a support
spider.

\begin{figure}[hbtp]
                                             \begin{center}
                                              \includegraphics[width=14cm]{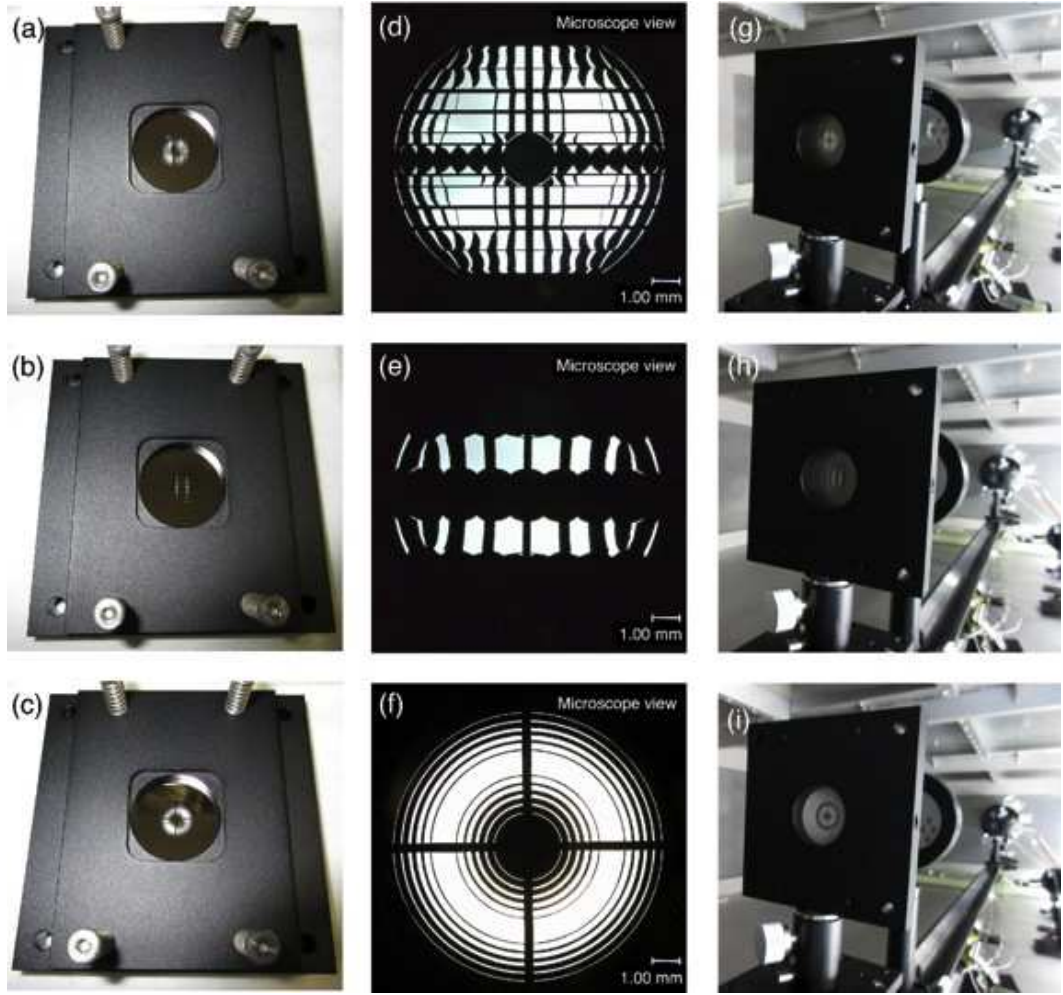}
                                             \end{center}
                                             \caption{ Panels (a), (d), and (g) are pictures of Mask-A. 
                                             Panels (b), (e), and (h) are pictures of Mask-B. Panels (c), (f), and (i) are pictures of Mask-C. 
                                             We adopted a 30 mm square design with a thicker handling area around the patterned region. 
                                             The size of the patterned region is 10 mm. 
                                             The thicknesses of the patterned and handling parts in the design are 5 $\micron$ and  $\geq$100 $\micron$, respectively. 
                                             }
                                             \label{fig2}
                                            \end{figure}

The theoretical PSFs of each mask are shown on the
right-hand side of figure \ref{fig1}. The central bright region of the
PSF is called the ``core", and the regions near to the core,
in which diffracted light is reduced, are called the ``Dark Regions (DRs)". 
Mask-A is an example of a design with
generalized darkness constraints. The inner working angle
($IWA$) and the outer working angle ($OWA$) are 3.3 $\lambda/D$
and 12 $\lambda/D$, respectively, where $\lambda$ is the wavelength and 
$D$ is the pupil diameter. The required contrast for this is
$10^{-5}$ at 3.3 $\lambda/D$ and $10^{-7}$ between 8 $\lambda/D$ and 12 $\lambda/D$. 
The required contrast between 3.3 $\lambda/D$ and 8 $\lambda/D$ was
determined by the constraint that the contrast should be
below a straight line on a log scale. Mask-B is intended
to be used with a small $IWA$. The required contrast for
this is $10^{-4}$ between 1.7 $\lambda/D$ and 6.2 $\lambda/D$. This mask is
more useful than the other two masks for direct observation
in the infrared wavelength region of young Jovian planets
close to the star. With Mask-C a wide-field coronagraphic
image is obtained. The $IWA$ and the $OWA$ are 5 $\lambda/D$, and
25 $\lambda/D$, respectively, and the contrast for this is $10^{-4.5}$ at
5 $\lambda/D$ and $10^{-7}$ between 12 $\lambda/D$ and 25 $\lambda/D$. The contrast
between 5 λ/D and 12 λ/D was determined to be below
a straight line on a log scale. This mask is useful for surveying
unknown exoplanets far from the stars they orbit 
and for observations of diffuse targets such as circumstellar
disks related to planetary formation in the infrared wavelength
region. The masks were fabricated as free-standing
masks, as shown in figure \ref{fig2}, and, consequently, can be
used for infrared observations. These free-standing masks
were fabricated in nickel using nano-fabrication technology
at HOWA Sangyo Co., Ltd and Photo Precision Co., Ltd
in Japan. The designs cover a 30 mm square, and comprise
10 mm mask patterns in thin nickel surrounded by
thicker nickel borders designed to enable the masks to be
easily handled. The target thicknesses of the patterned area
and the handling area are 5 $\micron$ and $\geq$ 100 $\micron$, respectively.
First, nickel with a target thickness of 5 $\micron$ was grown
by electrolytic plating on a temporary substrate. Then, the
handling area with a target thickness of $\geq$ 100 $\micron$ was
deposited by further electrolytic nickel plating. Finally, the 
free-standing mask was completed by stripping it from the
temporary substrate. The fabrication process for the freestanding
nickel masks is detailed in \cite{Enya2012}.

The pattern of the basic design of Mask-A contained
some isolated and/or ultra-fine features, which are not
viable for a free-standing mask. Therefore, for Mask-A, we
added bridge structures in a direction that had no effect on
the contrast to support and link the isolated and/or ultrafine
structures together. The fact that our mask design is
based on a 1D coronagraph is an important advantage in
that such bridge structures can in principle be applied \citep{Enya2011b}. 
Indeed, simulation showed that the contrast
obtained with the amended masks was equivalent to the
contrast obtained with masks with the basic design. On the
other hand, Mask-C has long arches and fine features not
shared by Mask-A and Mask-B (the width of the narrowest
arch and the space between the arches were designed to
be 33 $\micron$ and 20 $\micron$, respectively). Because of this, we
had the problem that the mask was irreversibly damaged
during removal from the substrate. However, the problem
was finally solved by small changes to the fabrication process
conditions for Mask-C (i.e., the mask design was not
changed) from the original fabrication process described in
\cite{Enya2012}. 

\subsection{Configuration of experiment}
Figure \ref{fig3} shows the instrument used for this work. All the
experimental optics were located in a clean-room at the
Japan Aerospace Exploration Agency/the Institute of Space
and Astronautical Science (JAXA/ISAS). The coronagraph
optics were set up in a vacuum chamber, an experimental
platform we call the High dynamic range Optical Coronagraph
Testbed (HOCT). We used a Super luminescent
Light Emitting Diode (SLED) with a center wavelength
of 650nm and wavelength width of 8 nm for the light
source. Light was passed into the chamber through a singlemode
optical fiber. The beam from the optical fiber was
collimated by a 50 mm diameter BK7 plano-convex lens
(SIGMA KOKI Co., Ltd.), and the collimated beam passing
through the pupil mask was focused by a second planoconvex
lens. The pupil mask was set at an angle of $\theta_{x} = 7^\circ$
to the plane perpendicular to the optical axis to remove
light reflected from the mask (see figure \ref{fig3} for the definition
of the coordinates and $\theta_{x}$). We used 3.4× relay optics
after the focal plane. Multi-layer anti-reflection coatings
optimized for wavelengths of 400 - 700 $\micron$ were applied to
both sides of the lens to reduce reflection. Though active
wavefront control helps to improve contrast, it was not
applied in this work in order to evaluate the performance
of the masks themselves. A commercially available cooled
CCD camera (BJ-42L, BITRAN) with 2048$\times$2048 pixels
set up in the chamber was used to measure the PSF. The
CCD was cooled and stabilized at $271.0\pm 0.5K (1\sigma)$.
This experimental system is capable of achieving a raw
contrast of $10^{-7}$ \citep{Haze2009, Haze2011, Haze2012, Enya2012}.

To obtain a high-contrast image, we carried out the following
procedure: we measured the core and the DR, each
of which have different imaging times, separately.When the 
DR was measured, we obscured the light from the core with
a focal plane mask (i.e., a DR-shaped hole mask) inserted at
the first focal plane after the pupil mask. For measuring the
core, we replaced the DR-shaped hole mask with two neutral
density (ND) filters. The transmission through the ND
filters is wavelength dependent, and at 650nm is 0.016\%.

\subsection{Core image}
The core image of the coronagraphic PSF was obtained
with exposure times of 0.3 s and 3 s. We inserted two ND
filters as previously mentioned. After each imaging process,
the laser source was turned off and a ``dark frame" measurement
was taken with the same exposure time and the
same neutral density filters. A ``raw" coronagraphic image
was obtained by subtracting the dark frame from the image
with the laser light on (see left-hand side of figure \ref{fig4}). These
results are quite consistent with those expected from theory
(see figure \ref{fig1}).

\subsection{DR image}
The DR of the coronagraphic image was observed with
exposure times of 0.3 s, 3 s, and 30 s. A ``dark frame" was
taken with the same exposure times, and these were then
subtracted from the DR images with the laser light on.
The observed DRs of the raw coronagraphic image, which
is the area of the image through the focal plane mask,
are shown on the right-hand side of figure \ref{fig4}. The contrast
was obtained by normalizing the observed DRs to the
peak of the core.

As shown in figure \ref{fig5}a, contrasts of $\sim10^{-4}$ - $10^{-7}$ and
$\sim10^{-7}$ for the ranges of 3.3 - 8$\lambda/D$ and 8 - 12$\lambda/D$, respectively,
were achieved using Mask-A. These experimental
contrasts almost reached the designed values. The slightly
poorer contrast near the $OWA$ is thought to be due to
contamination from light outside the $OWA$. As shown in
figure \ref{fig5}b, a contrast of $\sim10^{-4}$ close to the center was
achieved using Mask-B. The experimental contrast is almost
the same as the designed value. As shown in figure \ref{fig5}c, a
contrast of $\sim10^{-5}$ - $10^{-6}$ over an extended field of view
(5 - 25 $\lambda/D$) was achieved using Mask-C. We also found
speckle patterns other than diffraction patterns in the DRs.
For example, the observed PSF of Mask-C, which has a
contrast of $\sim10^{-6}$ between 20 $\lambda/D$ and 23 $\lambda/D$, has two
ring structures (see figures \ref{fig4}f and \ref{fig5}c).

\section{Discussion}
\subsection{Merits of the new masks for actual observations}
One of the critical issues for coronagraphy is preventing
the pupil of the telescope from being obscured. So offaxis
telescopes have been specifically designed for space
missions specializing in coronagraphy—e.g., TPF-C \citep{Traubetal}; SEE-COAST \citep{Schneider}; PECO
\citep{Guyon}. On the other hand, the masks developed
in this study are designed for an on-axis telescope,
in which the pupil is partially obscured by the secondary
mirror and its supporting arms. The experiments with each
of these masks described here show that the contrast is
significantly improved compared with non-coronagraphic
optics. These masks also have the properties of a binary
pupil mask coronagraph in that they are robust against
pointing errors and they can be used to make observations
over a wide wavelength range. These masks have no limitation
to the wavelength range over which they can be used
because they are free-standing. Therefore, the application
of these binary pupil mask coronagraphs to a number of
different normal, centrally obscured telescopes can help to
open up new platforms for coronagraphy. Many advanced
ground-based telescopes (e.g., current 8 - 10m class telescopes
like Subaru, and larger future ones such as TMT, 
E-ELT) and space telescopes (e.g., SPICA, JWST) with
an obscured pupil have the potential to be platforms for
coronagraphy over a wide wavelength region.
The specifications of the masks developed in this study
are complementary, and are, therefore, useful for observing
Jovian planets located at various distances from the central
star in the mid-IR wavelength region \citep{Fukagawa, Matsuo2011, Enya2011a}. Obtaining the
spectra of Jovian planets is invaluable if we wish to learn
more about planetary formation processes. These complementary
masks are also useful for studying protoplanetary
disks and AGN.

\subsection{The contrast obtained with Mask-C}
Here, we consider the factors that limited the contrast
obtained with the masks in the coronagraphic experiments.
The experimental PSF using Mask-B agreed very well with
the designed PSF. The experimental PSF using Mask-A
matched the designed PSF most closely. On the other hand,
forMask-C, the discrepancy in contrast between the design
and the experiment is significantly larger than for both
Mask-A and Mask-B. Moreover, the experimental PSF
obtained withMask-C had two ring structures not expected
from the design.

\begin{figure*}[hbtp]
                                             \begin{center}
                                              \includegraphics[width=14cm]{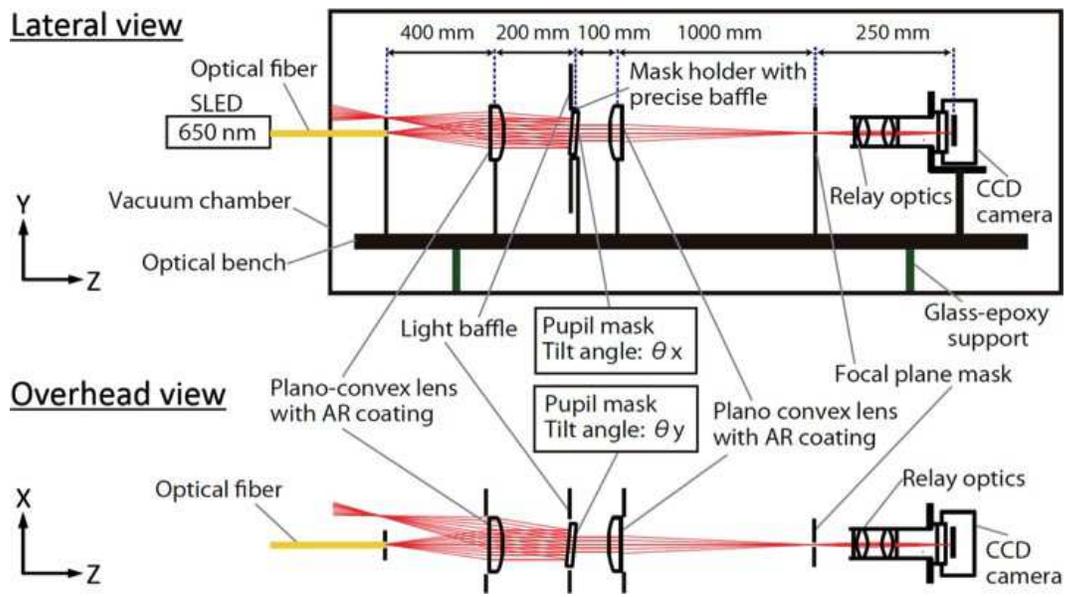}
                                             \end{center}
                                             \caption{ Lateral and overhead views of the configuration of the experimental optics.}
                                             \label{fig3}
                                            \end{figure*}

\begin{figure}[hbtp]
                                             \begin{center}
                                              \includegraphics[width=12cm]{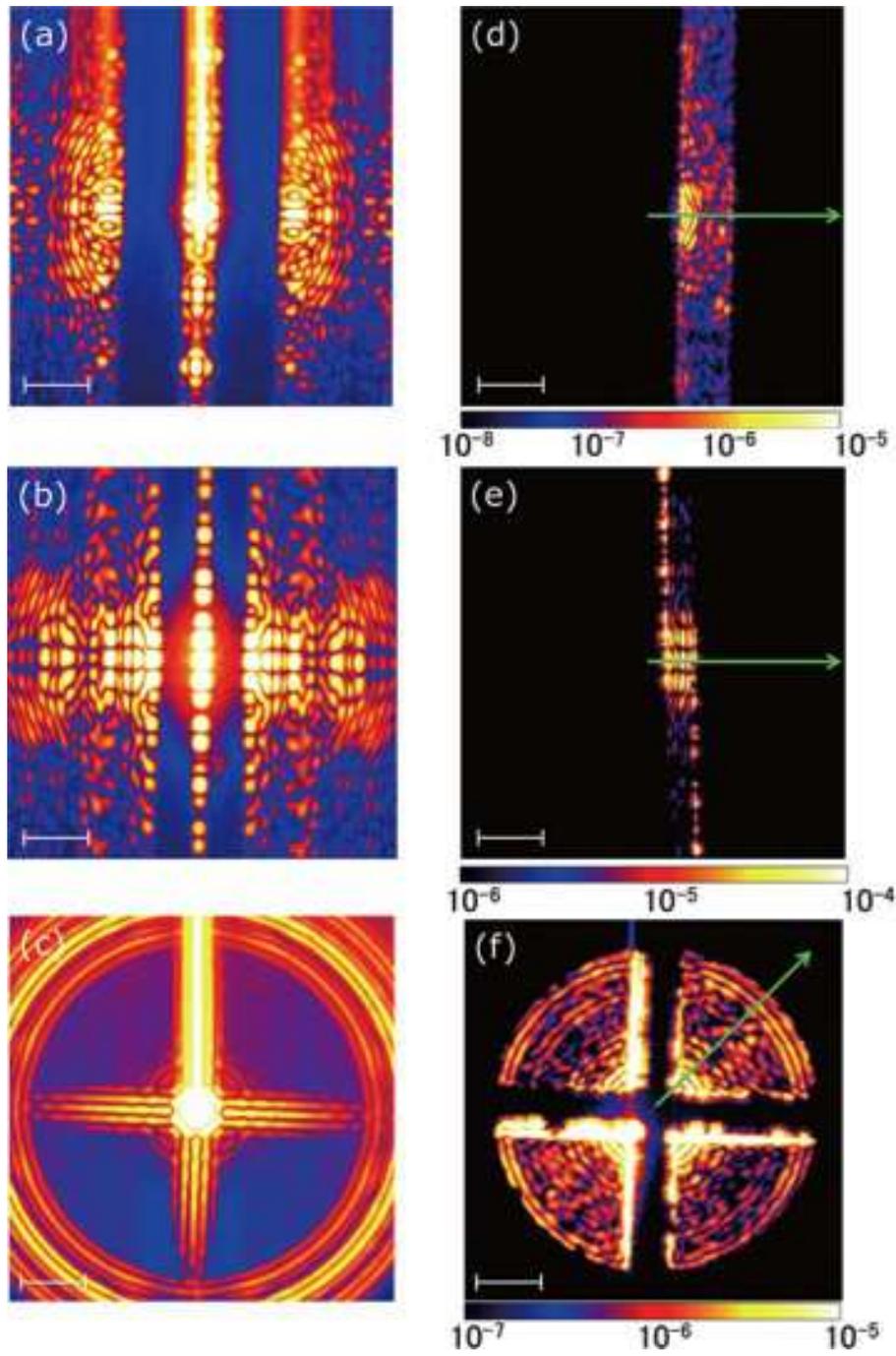}
                                             \end{center}
                                             \caption{Observed coronagraphic images obtained with a SLED (650 nm). 
                                               Panels (a), (b), and (c) show the observed coronagraphic PSFs of Mask-A, Mask-B and Mask-C, respectively. 
                                               Panels (d), (e), and (f) show the DR images obtained with the focal plane mask. 
                                               The green arrow indicates the direction of the radial profile. 
                                               The scale bars correspond to 10$\lambda/D$. 
                                               }
                                             \label{fig4}
                                            \end{figure}

\begin{figure}[hbtp]
                       \begin{center}
                        \includegraphics[width=6cm]{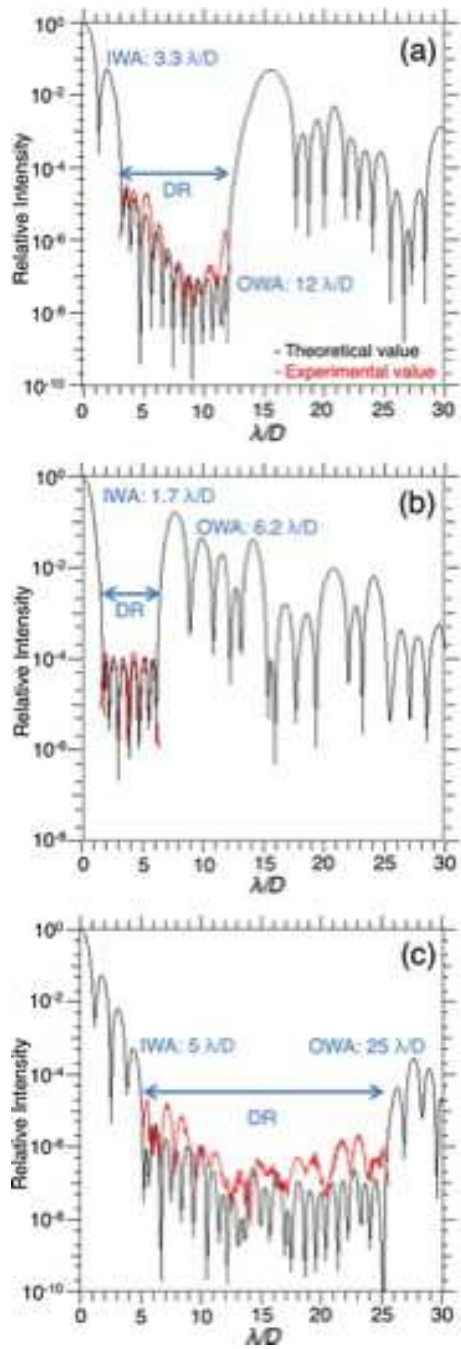}
                       \end{center}
                       \caption{Radial profiles of the observed (red line) and the theoretical (black line) coronagraphic PSF.
                       Each profile is normalized by the peak intensity. 
                       Panel (a): PSF profiles of Mask-A. $IWA$ is 3.3$\lambda/D$. $OWA$ is 12$\lambda/D$. 
                       Panel (b): PSF profiles of Mask-B. $IWA$ is 1.7$\lambda/D$. $OWA$ is 6.2$\lambda/D$.
                       Panel (c): PSF profiles of Mask-C. $IWA$ is 5$\lambda/D$. $OWA$ is 25$\lambda/D$.
                       }
                       \label{fig5}
                      \end{figure}

In the experiment described in the previous section, the
pupilmaskwas set at an angle ($\theta_{x} = 7^\circ$, using the coordinate
system defined in figure \ref{fig3}) to the plane perpendicular to
the optical axis in order to eliminate light reflected by the
mask. Thus, to enable further discussion of the contrast
obtained from the experiment, we examined the effect of
tilting the mask.

Using Cartesian coordinates, the x and y variables for
the 1D coronagraph masks, Mask-A and Mask-B, can
be separated, while those for Mask-C, which is rotationally
symmetric, cannot. Qualitatively, this suggests that
the contrast can be reduced from the optimum by tilting
Mask-C. However, a quantitative assessment of the influence
of the tilt on the contrast obtained with Mask-C has
not yet been done.

\begin{figure}[hbtp]
                       \begin{center}
                        \includegraphics[width=14cm]{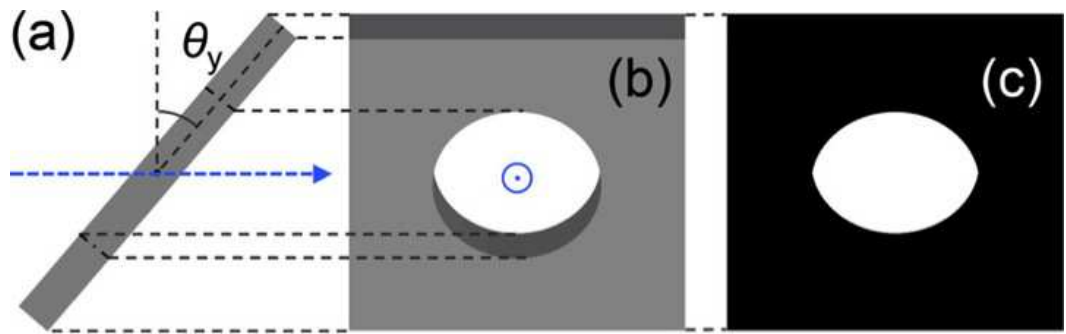}
                       \end{center}
                       \caption{Conceptual diagram of the mask for use in simulating a tilted pupil mask. 
                       The blue arrow indicates the direction of the optical axis. 
                       (a) Overhead view of the mask rotated by $\theta_{y}$ from a plane perpendicular to the optical axis.
                       (b) Lateral view of the mask rotated by $\theta_{y}$ from a plane perpendicular to the optical axis.
                       The dark gray region shows the mask thickness of 5 $\micron$. 
                       (c) Projection of the tilted mask (b) on to a perpendicular plane including the influence of the mask thickness of 5 $\micron$. 
                       The transmission through the black and white regions is 0 and 1, respectively. 
                       The PSF is simulated by using the Fourier transform of the pupil. 
                       }
                       \label{fig6}
                      \end{figure}

\begin{figure*}[hbtp]
                       \begin{center}
                        \includegraphics[width=16cm]{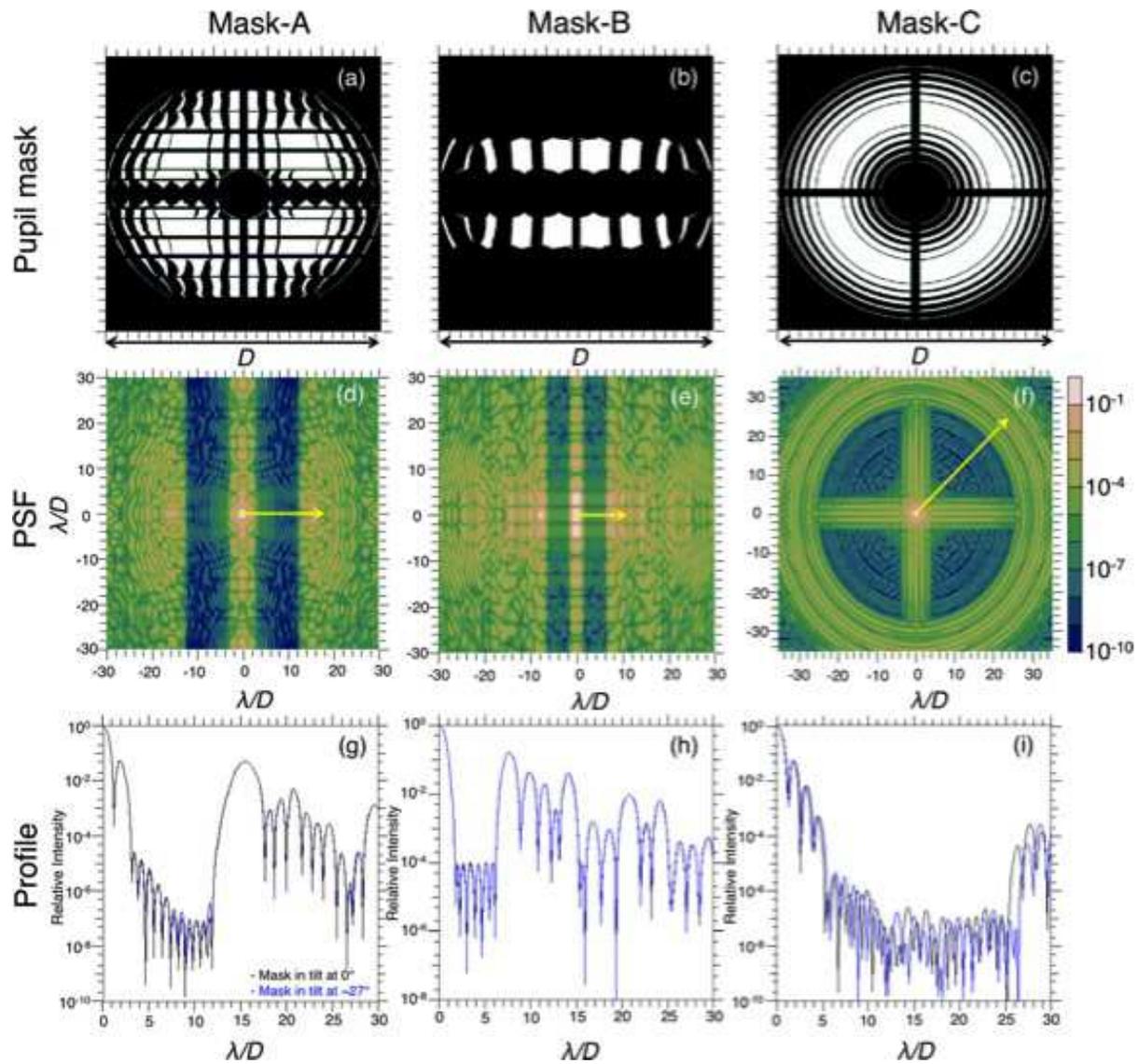}
                       \end{center}
                       \caption{ 
                       Results of simulation with the pupil mask tilted (Pupil mask, PSF and Profile). 
                       Panels (a), (b), and (c) show simulated projections for the tilted masks (Mask-A, Mask-B, and Mask-C all tilted at $\sim27^\circ$). 
                       Panels (d), (e) and (f) show the theoretical coronagraphic PSFs from (a), (b) and (c), respectively.
                       The yellow arrow indicates the direction of the radial profile. 
                       Panels (g), (h) and (i) show the PSF profiles when using the masks tilted at $0^\circ$ (black line) and at $\sim27^\circ$ (blue line).
                       There is no reduction in contrast in each case. 
                       }
                       \label{fig7}
                      \end{figure*}

Therefore, we assumed the masks to be ideally flat and,
with a larger tilt angle ($\theta_{y} = 27^\circ$, using the coordinate
system defined in figure \ref{fig3}), made projections that took
account of the influence of the mask thickness of 5 $\micron$, as
shown in figure \ref{fig6}. The coronagraphic PSFs were obtained
from simulation using Fourier transform techniques. The
results are shown in figure \ref{fig7}. The simulation shows that a
tilt angle of $27^\circ$ should not affect the contrast obtained not
only with Mask-A and Mask-B but also with Mask-C.

\begin{figure}[hbtp]
                       \begin{center}
                        \includegraphics[width=6cm]{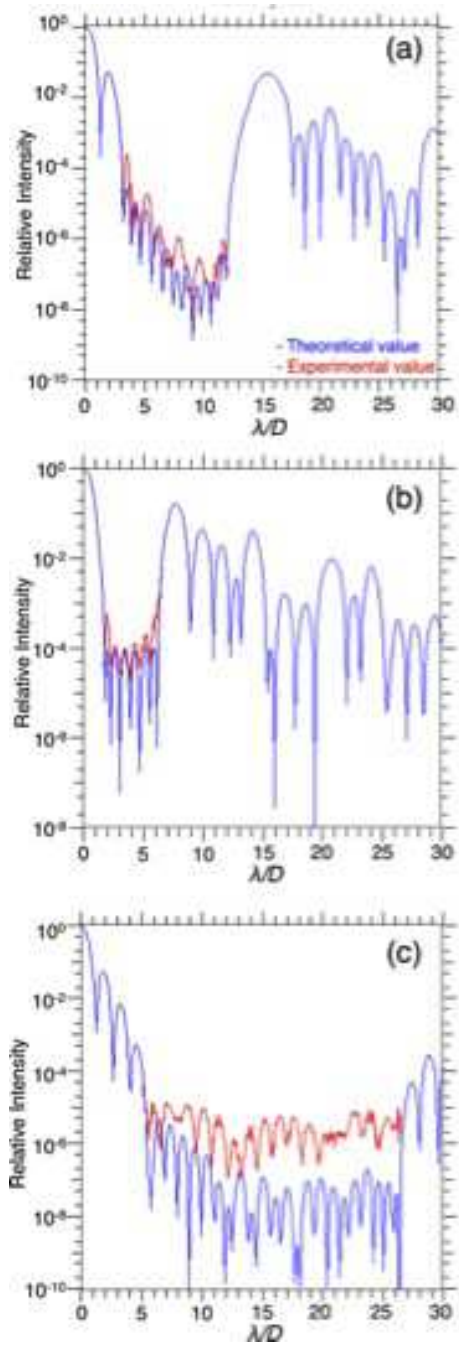}
                       \end{center}
                       \caption{Radial profiles of the observed (red line) and the theoretical (blue line) coronagraphic PSFs with the mask tilted at $\sim27^\circ$. 
                       Panels (a), (b), and (c) show the PSF profiles of Mask-A, Mask-B, and Mask-C, respectively.
                       }
                       \label{fig8}
                      \end{figure}

Consequently, we conducted additional experiments
with the masks tilted at $\theta_{y} = 27^\circ$ (see figure \ref{fig3}). 
A comparison between the results from experiment and those from
simulation with the masks tilted at $27^\circ$ is shown in figure \ref{fig8}.
With Mask-A and Mask-B the experimental results are
almost consistent with the simulations. The slightly poorer
contrast near the $IWA$ and the $OWA$ is thought to be due to
the effect of light contamination from outside the $IWA$ and
the $OWA$.On the other hand, the experimental results using
Mask-C show the contrast to be worse than that expected
from the simulation. A deviation from perfect flatness of
the mask is a possible reason for these results. Our freestanding
masks were fabricated by stripping them from a
substrate after forming them on the substrate. Indeed,many
trials were needed to succeed in stripping off the long arches
and fine structure of Mask-C without it breaking because 
the stripping process for Mask-C is fraught with more difficulty
than for either Mask-A or Mask-B. To assess and
possibly prevent the effects due to errors in flatness, comparative
experiments with a similar mask on a substrate
would be useful.

\section{Conclusion}
In this study, we have presented the fabrication and experimental
demonstrations of free-standing coronagraphmasks
with three complementary designs for telescopes, in which
the pupil is partially obscured by a secondary mirror and
its support structure. Coronagraphic images obtained from
the experiments weremostly consistent with the design. The
masks can be tilted with little effect on contrast, thereby
giving us the freedom to choose the tilt angle to avoid
ghosting and to manage stray light. The results are important
for allowing general-purpose telescopes not specialized
for coronagraphy to be used as platforms for high contrast
coronagraphic observations in future.

 \section*{Acknowledgement}
First of all we are grateful to R. J. Vanderbei and the LOQO solver
presented by him \citep{Vanderbei1999}. This work was financially supported
by the Japan Science and Technology Agency and Grants-in-
Aid for Scientific Research (Nos. 24840049 and 22244016) from the
Japan Society for the Promotion of Science. We thank T. Ishii from
Photo Precision Co., Ltd., A. Suenaga from HOWA Sangyo Co., Ltd,
and their colleagues in each of these companies.

\newpage

%
\begin{table}[hbtp]
\begin{minipage}{\textwidth}
  \caption{Required contrast in the mask design.}
  \label{table1}
  \centering
  \begin{tabular}{lcr}
    \hline
    Mask (type)  & Separation angle \footnotemark[$\ast$]  &  Contrast  \\
                         & (λ/D)             &                  \\
    \hline 
    Mask-A (integral 1D)  & 3.3 \footnotemark[$\dagger$]  & $10^{-5}$ \\
                                       & 8.0   & $10^{-7}$ \\
                                       & 12  \footnotemark[$\ddagger$]  & $10^{-7}$ \\
    Mask-B (integral 1D)  & 1.7  \footnotemark[$\dagger$]  & $10^{-4}$ \\
                                       & 6.2  \footnotemark[$\ddagger$] & $10^{-4}$  \\
    Mask-C (ring)             & 5.0  \footnotemark[$\dagger$]  & $10^{-4.5}$ \\
                                       & 12    & $10^{-7}$ \\
                                       & 25  \footnotemark[$\ddagger$]  & $10^{-7}$ \\

    \hline
  \end{tabular}
\footnotetext[$\ast$]{Separation angle for integral 1D mask is defined from the PSF core to the coronagraphic direction.}
\footnotetext[$\dagger$]{$IWA$ of the mask design (see text).}
\footnotetext[$\ddagger$]{$OWA$ of the mask design (see text).}
\end{minipage}
\end{table}
 %

\end{document}